\documentclass[mathleft
]{an}
\usepackage{graphicx}
\usepackage{times}
\usepackage{aasmac}
\usepackage{amssymb}
\newcommand{\mrn}{\mathrm {Rm}}

\usepackage{natbib}
\bibpunct{(}{)}{;}{a}{}{,}
\def\gsim{\lower.4ex\hbox{$\;\buildrel >\over{\scriptstyle\sim}\;$}} 
\def\lsim{\lower.4ex\hbox{$\;\buildrel <\over{\scriptstyle\sim}\;$}} 
\begin{document}

\Pagespan{1}{}
\Yearpublication{2012}%
\Yearsubmission{2012}%
\Month{3}%
\Volume{XXX}%
\Issue{X}%

\title{Linear stability analysis of the Hall magnetorotational instability in a spherical domain}

\author{T. Kondi\'{c}\inst{}\thanks{Corresponding author:
  {T.Kondic@leeds.ac.uk}}
\and  G. R\"{u}diger\inst{} 
\and  R. Arlt\inst{}
}
\titlerunning{Linear stability analysis of the Hall-MRI in a spherical domain}
\authorrunning{T. Kondi\'{c}, G. R\"udiger \& R. Arlt}
\institute{
Leibniz-Institut f\"{u}r Astrophysik Potsdam, An der Sternwarte 16, 
D-14482 Potsdam, Germany}

\received{XXX}
\accepted{XXX}
\publonline{XXX}


\abstract{We investigate the stability of the Hall-MHD system and
determine its importance for neutron stars at their birth, when they
still consist of differentially rotating plasma permeated by extremely
strong magnetic fields.  We solve the linearised Hall-MHD equations in a
spherical shell threaded by a homogeneous magnetic field. With the
fluid/flow coupling and the Hall effect included, the magnetorotational
instability and the Hall effect are both acting together. Results differ
for magnetic fields aligned with the rotation axis and anti-parallel
magnetic fields. For a positive alignment of the magnetic field 
the instability grows on a rotational time-scale for any 
sufficiently large magnetic Reynolds number. Even the magnetic 
fields which are stable against the MRI due to the magnetic
diffusion are now susceptible to the shear-Hall instability.  In
contrast, the negative alignment places strong restrictions on the
growth and the magnitude of the fields, hindering the effectiveness of
the Hall-MRI. While non-axisymmetric modes of the MRI can be suppressed
by strong enough rotation, there is no such restriction when the Hall effect
is present. The implications for the magnitude and the topology of
the magnetic field of a young neutron star may be significant.  }

\maketitle

\section{Introduction}

Neutron stars are usually formed as a result of the gravitational
collapse of a star at least eight times more massive than the Sun.
Degenerate electron pressure, diminishing as the most energetic
electrons begin interacting with the protons in the nuclei producing
neutrons and neutrinos, is unable to stop the gravitational collapse
until the nuclear densities are reached \citep{shte1983}. 

The transition from the collapsing stellar core to a stable neutron star
occurs through the protoneutron star stage. At the end of the collapse,
the remnant stabilises into a hot, superdense, degenerate nuclear fluid
confined beneath the bounce shock zone of the ensuing supernova
\citep{pagewe2006} . The proto-neutron star (PNS) has initially a
diameter of around $100\ \mbox{km}$ which reduces to the canonical $20\
\mbox{km}$ in one second \citep{bula1986}. The final transformation of
the core remnant into a young neutron star lasts for additional one hundred
seconds, during which it settles in the state of cold catalysed matter
\citep{bourbe2006}.

Simulations of the core collapse show that almost any initial rotation
profile, including rigid rotation, results in a differentially rotating
PNS \citep{akwh2003,obal2006,otbuth2006}. Depending on the initial spin
of the progenitor core, the collapse may result in a strong differential
rotation with a negative radial gradient. Such conditions are favourable
for the occurrence of the magnetorotational instability (MRI)
\citep{baha1991}.

Because of the very short time-frame during which hydrodynamical
instabilities can take place in the PNS, the exponential growth of the
magnetic field provided by the MRI may be an essential ingredient in
bridging the gap between the relatively weak progenitor magnetic fields
and those observed in strongly magnetised neutron stars. It may also
play an important role in the mechanism which powers supernova
explosions \citep{akwh2003,arbi2005,obal2006,bude2007,obce2009}.

The magnitudes of the magnetic field during the PNS stage of up to $10^{15}\
\mathrm{G}$ raise the question of how important the Hall effect may
be for its evolution. So far, the theory of the interaction between the
MRI and the Hall effect has mostly been flourishing in the context of
weakly ionised accretion disks.

The local analysis of the Hall-MHD in differentially rotating disks
\citep{wa1999,bate2001,urrue2005} revealed that the influence can be
both stabilising or destabilising, depending on a number of factors,
including the orientation of the magnetic field. \citet{rueki2005} found
similar results concerning the global instability in protostellar disks
and determined threshold magnetic fields for which the Hall induced
instability would occur. 

An interesting feature observed in the these works was that an
instability occurs even when the conditions for the usual MRI are not
fulfilled. For fields parallel to rotation axis, weaker than expected
magnetic fields are susceptible to the instability. In addition, even a
positive differential rotation gradient -- stable against the MRI -- can be
destabilised. This is an indication that there are new channels through
which the magnetic field taps into the energy of the shear when the Hall
effect is present. This is even more obvious in models with pure shear 
without rotation \citep{ku2008,bego2011}. 

\citet{korue2011} examined the instability driven by the Hall effect
itself, the shear-Hall instability (SHI), in the context of protoneutron
stars. Like the MRI, it feeds off the shear energy and grows on the
time-scale of the rotation period. They show that the SHI may, if the
conditions are fulfilled, occur in the PNS, but they focus solely on
solving the induction equation, thus leaving out the back-reaction of
the field on the flow.

Unlike \citet{korue2011}, this works includes the back-reaction leading
to the interplay between the Hall effect and MRI. Stability maps and
growth rates of the global Hall-MRI instability follow from numerical
calculations of the Hall-MHD equations in a spherical shell. These
results are compared to the properties of a protoneutron star.

\section{Model}

The coupling between the magnetic field and the flow responsible for the
MRI and the SHI is contained in the Navier-Stokes and the induction
equations of the MHD system. In the incompressible and isothermal case, these
read
\begin{equation}
\frac{\partial\vec{u}}{\partial t}+(\vec{u}\cdot\nabla)\vec{u} =  -\nabla p + \frac{1}{4\pi\rho}(\nabla\times\vec{B})\times\vec{B}+\nu\Delta\vec{u},
\label{eq:ns}
\end{equation}
\begin{eqnarray}
\lefteqn{\frac{\partial\vec{B}}{\partial t}  =  
-\nabla\times\left(\frac{c^{2}}{4\pi\sigma}{R_B}(\nabla\times\vec{B})\times\vec{e}_B\right) -} \nonumber \\
& & \quad\quad \nabla\times\left(\frac{c^{2}}{4\pi\sigma}\nabla\times\vec{B}\right)+ \nabla\times(\vec{u}\times\vec{B}).
\label{eq:induction}
\end{eqnarray}
The variables $\vec{u}$ and $\vec{B}$ are, respectively, the velocity and 
the magnetic field, while $p$ is the pressure. The Hall parameter 
${R_{B}}$ is the ratio of the off-diagonal to the transverse
component of the conductivity tensor and scales linearly with the
magnetic field. The viscosity is represented by $\nu$, mass density by
$\rho$ and the isotropic diagonal element of the electric conductivity 
tensor by $\sigma$. The unit vector $\vec{e}_B$ gives the orientation 
of the magnetic field.

Equations (\ref{eq:ns}) and (\ref{eq:induction}) do not differ from
those used in many other works dealing with weakly ionised plasmas.
However, the physics behind the interactions of different plasma species
is not the same.  The two charged components considered implicitly
in this model are fully ionised atoms and electrons.  The Eq.~(\ref{eq:induction}) is what remains after all the terms of order
$m_{\mathrm e}/m_{\mathrm i}$ are neglected, where $m_{\rm e,i}$ 
are the masses of electrons and ions. This can be interpreted as if the
motion of the fluid is entirely due to the ions, while the current
carriers are the much lighter electrons. If the magnetic field is weak,
or the collisional frequency between the species is large,
${R_{B}}$ tends to zero and ordinary MHD is recovered. The Hall
effect is then a correction to the ordinary MHD which arises from a
response of electrons to the magnetic field.

Because the aim of this work is the investigation of the conditions
under which the instability occurs, the system can be linearised around
a stable (or a very slowly evolving) state. We chose a spherical shell
with radial boundaries at $r_{\rm i} = 0.7$ and $r_{\rm o}=1$. The shear necessary for the
operation of the MRI and the SHI is provided by a hydrodynamically stable 
differential rotation profile
\begin{equation}
\Omega(s) = \Omega_{0} \left[1+(s/s_{0})^2\right]^{-1/2},
\end{equation}
where $\Omega_{0}$ is the angular velocity
on the rotation axis, $s$ the cylinder distance from it and $s_0=0.5$. 
The spherical components of the background
velocity, $u_r$, $u_\theta$, $u_\phi$ are thus
$\vec{u_{0}}(s)=(0, 0, s \Omega)$.
A homogeneous magnetic field $B_{0}\mathrm{ \vec{e_{z}}}$ is imposed in the
direction of the rotation axis. This choice prevents a coupling
between the flow and the field in the absence of a perturbation.

The domain of linear solutions is restricted to a spherical shell of
constant mass density interfacing to the current-free surroundings on
both of its sides.  This choice was made for the sake of numerical
tractability of the problem, even though this is a considerable
simplification for the inner boundary. 

It is now possible to write the linear form of the equations. With the separation of the magnetic field into
 perturbations and background field, $\vec{B}=\vec{b}+\vec{B}_0$, we can write 
\begin{eqnarray}
\lefteqn{\frac{\partial\vec{u}}{\partial t}   = 
{S}^{2}(\nabla\times\vec{b})\times\vec{e}_{z}\ - \nabla f +}  \nonumber \\ 
				&& \quad\quad \vec{u}_0\times \nabla\times \vec{u}+\vec{u}\times\nabla\times\vec{u}_0+\mathrm{Pm}\Delta\vec{u}, 
\label{eq:linflow}
\end{eqnarray}
\begin{eqnarray}
\lefteqn{\frac{\partial\vec{b}}{\partial t}  =  -R_{B}\nabla\times \left[(\nabla\times\vec{b})\times\vec{e}_{z}\right]- } \nonumber \\
& &\quad\quad \nabla\times(\nabla\times\vec{b})
+\nabla\times(\vec{u}_0\times\vec{b}+\vec{u}\times \vec{B}_0) ,
\label{eq:linind}
\end{eqnarray}
where all gradient terms are collected in $\nabla f$, which will vanish as
we are taking the curl of Eq.~(\ref{eq:linflow}) for the solution. The time is normalised to the diffusion time-scale
$\tau_{\mathrm{D}}=R^2/\eta,\ \eta=c^2/(4\pi\sigma)$
and all the units of length to the radius of the outer boundary
$R$ of the star in Eqs.~(\ref{eq:linflow}) and (\ref{eq:linind}).
The Lundquist number is the ratio between the diffusion time and the
Alfv\'{e}n time, $\tau_{ {\mathrm A}}=\sqrt{4\pi\rho}\,R/{B_{0}}$), whence
\begin{equation}
  S = \frac{RB_0}{\sqrt{4\pi\rho} \, \eta}.
\end{equation}
The magnetic Prandtl number $\mathrm{Pm}$ is the ratio of
viscosity to magnetic diffusivity and is set to unity in 
the computations presented in this paper. The magnetic
Reynolds number is ${\mathrm{Rm}=R^2\Omega_{0}/\eta}$.  Finally,
$R_{B}$ denotes the Hall parameter fixed by the magnetic field
$B_{0}$. As the density is taken to be constant, these parameters do not
vary in the domain.

The information about the conditions for the onset of the instability
and the time-scale of its development is contained in the growth rate of
the eigenvalue solution of the 
system (\ref{eq:linflow})--(\ref{eq:linind}).  A linearised version of the pseudo-spectral, spherical MHD code 
by \citet{ho2000} was used to determine the growth rates and the stability 
criteria.

The results of a local instability analysis are useful to determine the
numerical limitations of the global approach. According to the ideal MRI
dispersion relation \citep{baha1998}, the length-scale of the fastest
growing MRI mode, for a given magnetic field tends to zero as the shear
becomes weaker.  An ideal numerical simulation would face a lower limit of the shear of showing instability 
because of the non-vanishing numerical diffusivity. In our case, the presence of a physical diffusivity sets
 a lower limit on the shear (as well as on the magnetic field for given shear) to be MRI unstable.
 We can compare the growth rate of the instabilities with the decay rate of the diffusion for a
 given wavenumber $k$, $\omega_{\rm d}=-\eta \, k^2$, and thus derive a maximum wavenumber $k_{\rm max}$
 which the unstable mode of the MRI should have to be excited at a given finite $\eta$. From
\begin{equation}
\omega=\left|\frac{1}{2} \frac{{\rm d}\,\Omega}{{\rm d}\, \ln \, s}\right|= -k_{\rm max}^2 \, \eta
\label{om}
\end{equation}
we obtain
\begin{equation}
k_{\rm max}= \frac{1}{2}\sqrt{\frac{\Omega_0}{\sqrt{2} \, \eta}}
\label{kmax}
\end{equation}
for the assumed rotation profile and ${s=s_0=1/2}$. In our dimensionless units, this corresponds to $k_{\rm max}=13$
at ${\rm Rm} = 1000$ and $k_{\rm max}=23$ at ${\rm Rm} = 3000$. 
The resolution of 20~Chebyshev polynomials and 70~Legendre 
polynomials was enough to resolve all the corresponding length scales
in the range of investigated values of ${\rm Rm}$ and $S$.  

\section{Results} 

The eigensolutions of the system (\ref{eq:linflow})--(\ref{eq:linind}) are
divided in separate classes defined by the transformation properties of the
equations. The system is invariant under both rotations around $\vec{e}_z$ and
reflection in the equatorial plane. The former implies on $\rm{e}^{im\phi}$
behaviour in the azimuthal direction, while the latter distinguishes between
antisymmetric field (accompanied by symmetric flow) and symmetric field
(antisymmetric flow) behaviour in the ($r$, $\theta$)-plane. The symmetric 
field has the property that only its $\theta$-component reverses
sign under symmetry transformation, while the radial and azimuthal
components remain the same. The opposite holds for the antisymmetric field.
The antisymmetric and symmetric ${m=0}$ solutions are labeled with A0 and S0,
respectively, in this manuscript. The label A0 thus means that $b_r(r, \theta, \phi) =
-b_r(r, \pi-\theta, \alpha\phi)$, $b_\theta(r, \theta, \phi) = b_\theta(r, \pi-\theta, \alpha\phi)$,
$b_\phi(r, \theta, \phi) = -b_\phi(r, \pi-\theta, \alpha\phi)$, while $u_r(r, \theta, \phi) =
u_r(r, \pi-\theta, \alpha\phi)$, $u_\theta(r, \theta, \phi) = -u_\theta(r, \pi-\theta, \alpha\phi)$,
$u_\phi(r, \theta, \phi) = u_\phi(r, \pi-\theta, \alpha\phi)$ for arbitrary $\alpha$.

We note that there is a difference between symmetries of the linearised and
the full induction equation. While the solutions of the linearised equation 
can be of either A or S symmetry types, only the A type is possible in the nonlinear
case (see \citealt{horu2002} for detailed explanation of the symmetries of the Hall
induction equation).  The absence of total equatorial symmetry has exciting 
consequences for the resulting topology of the magnetic field, 
as we shall explain in the final section.

\begin{figure}
	\includegraphics[clip,width=0.48\textwidth]{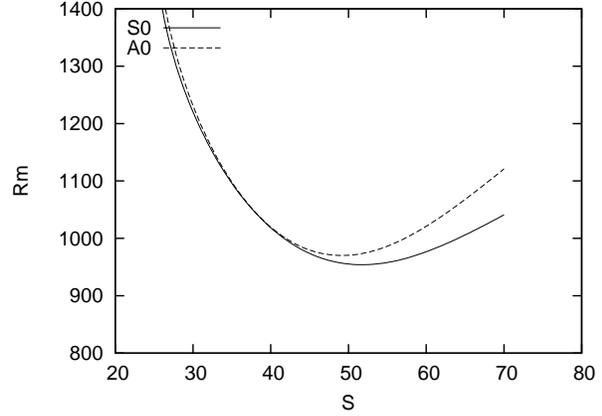} 
	\caption{Pure MRI stability curves for axisymmetric perturbations ($m=0$) which
        are symmetric (S0) and antisymmetric (A0) with respect to the equator.}\label{fig:staAS0} 
\par\end{figure}

\begin{figure}

	\includegraphics[clip,width=0.48\textwidth]{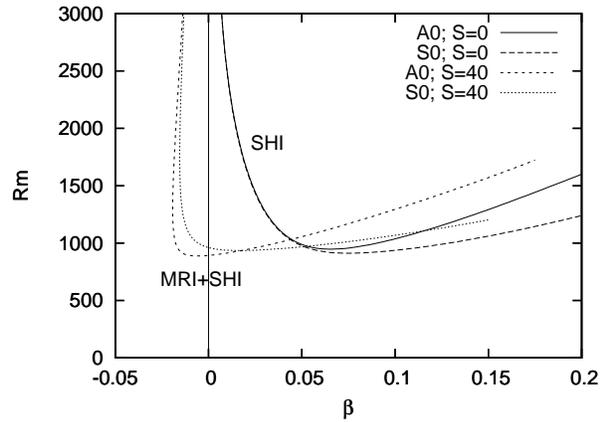}
	\caption{Hall-MRI stability curves for the S0 and A0 modes. The
	instability region is inside the curves. The influence of the
	Hall effect, for a fixed background field ($S={\rm const}$)
	is represented by $\beta=R_B/|S|$. Pure SHI corresponds to
	$S=0$ case.}\label{fig:staAS040} 
\par\end{figure}

\begin{figure}
	\includegraphics[clip,width=0.48\textwidth]{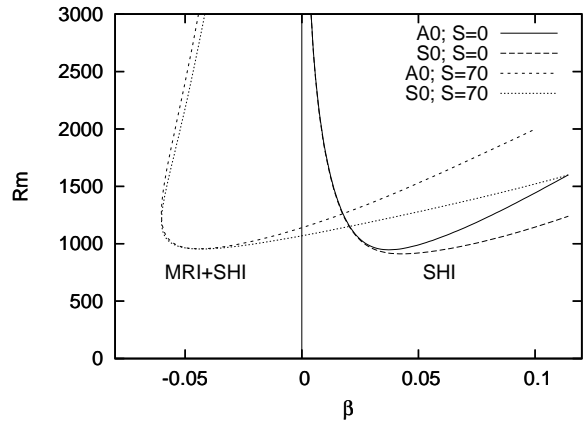} 

	\caption{Hall-MRI stability curves for S0 and A0 modes. The
	instability region is inside the curves. The influence of the
	Hall effect, for a fixed background field ($S={\rm const}$)
	is represented by $\beta=R_B/|S|$. The solid and long-dashed curves are solely due to the SHI.}\label{fig:staAS070} 
\par\end{figure}

\subsection{Stability maps} 

The curves in Figs. \ref{fig:staAS0}, \ref{fig:staAS040}, and
\ref{fig:staAS070} show the marginal stability of the Hall-MRI system.
In Figs. \ref{fig:staAS040} and \ref{fig:staAS070}, $\beta$ is the ratio
between the Hall parameter and the absolute value of the Lundquist
number, $\beta=R_B/|S|$. For non-quantising fields, it depends only on the sign of the
magnetic field and the structural properties of the plasma.  It can be
thought of as a measure of importance of the Hall effect for a fixed
background field.

The curves in Fig. \ref{fig:staAS0} represent the pure MRI. Note the
transition to stability for ${S\la25}$ due to diffusion. Here, the
influence of the field on the fluid is exerted through the Lorentz
force, but there is no transformation of toroidal into poloidal magnetic field since there is no Hall
effect.

The limit of small $S$ is the opposite case. For a very long
Alfv\'{e}n time, there is only amplification of magnetic perturbations by the differential rotation.  This
case is represented by the ${S=0}$ line in Figs. \ref{fig:staAS040} and
\ref{fig:staAS070}.  In the framework of ordinary MHD, this would be a
stable situation. As it can be seen, however, the Hall effect gives rise
to the SHI.  The shear feeds the energy into the toroidal field, and the
Hall effect channels it back to poloidal creating a positive feedback.
This is  the same as in \citet{korue2011}.  The consequence is that, due
to the shear-Hall mechanism, there is an instability even in the range $S=0$ to
$S=25$. 

Outside of the diffusion zone, the stability is under larger influence
of the interaction between Navier-Stokes and induction equation (positive $\beta$ in Figs.
\ref{fig:staAS040} and \ref{fig:staAS070}) than the shear-Hall
mechanism, with the critical $\mrn$ being closer to the pure MRI values.
For a certain window of $\beta$, the presence of the momentum equation has a slightly stabilizing effect compared to pure SHI in the induction equation only. 

The results for $\vec{B}_0$ anti-parallel to the rotation axis are more
dramatic. There is no pure SHI in this case, but the Hall effect still
produces some important differences. The instability zone  now also possesses 
an upper limit of Rm beyond which the system becomes stable.  On approaching 
the upper boundary, the length-scale of unstable perturbations increases, 
until it finally exceeds the domain size at the critical line. A similar 
effect  was noted in a local approximation by \citet{bate2001}.

The fact that the opposite orientation of $\vec{B}_0$ and $\vec{\Omega}$ is
less stable than when the two vectors are aligned, is already known
from accretion disk theory \citep{rueki2005}. In this case, the Hall
effect enhances the MRI mechanism. However, for large enough $|\beta|$
at  a given $S$ or for small enough $S$ at a given $\beta$,
the Hall effect suppresses the instability, rendering the system stable. 
Apparently, there is no way for the MRI to set in for substantially lower 
critical Rm with the help of the Hall effect.

The comparison of Figs. \ref{fig:staAS040} and \ref{fig:staAS070} also
shows how the ``depth'' (minimal unstable $\beta$) of the instability
zone increases, if the field grows to larger strengths (larger $|S|$).  

To summarise, if the field and the angular velocity vector are
anti-aligned, the Hall effect severely restricts the maximal magnetic
field that can be  MRI unstable. There is no such restriction for
the positive orientation.  This is a crucial result, calling for
reassessment of any conclusions about the field patterns emerging in
differentially rotating, strongly magnetised systems, derived assuming the
 validity of ordinary MHD approximation (for example,  a
body of work dealing with the MRI in PNS).

\subsection{Growth rates}
\begin{figure}
	\includegraphics[clip,width=0.48\textwidth]{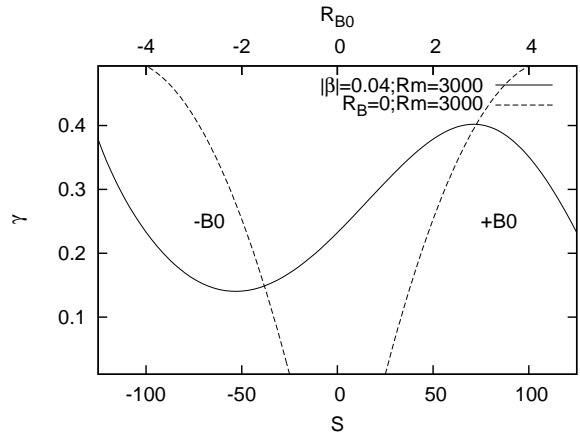}

	\caption{The relationship between the growth rate (normalised to
	$\Omega_{0}/(2\pi)$)  and the magnetic field. In this graph the
	ratio between the Hall parameter and the Lundquist number is
	kept constant.  Therefore, changing $S$ amounts to changing
	the strength of the $B_0$ field. The magnetic
	Reynolds number is kept fixed.} \label{fig:grA0fixb} 

\end{figure}

We varied the strength of the magnetic field but kept the ratio of Hall parameter to Lundquist number, $\beta$, fixed. The resulting growth rates are shown in Fig.~\ref{fig:grA0fixb}.

Similar to the findings by \citet{rueki2005}, the
Hall effect shifts the extrema of the graph with, again, evident
stabilising influence for the positive orientation (for $S\gsim 70$, $R_B>0$). Also the largest growth rate obtained for Hall-MRI is limited and smaller than the growth rates of the MRI for large enough $S$. For $S\lsim 100$, the growth rates of the Hall-MRI is  smaller for an anti-parallel orientation of the magnetic field and the rotation than for parallel orientation. Only for very strong anti-parallel magnetic fields, the growth rates exceed the parallel case, but never exceeds the growth rate of the pure MRI for $S\gg 1$.  The maximal growth
rate was of the order of the inverse rotation period, the
smallest time-scale of the system, similar to what was found by
\citet{korue2011} for the SHI above.

\begin{figure*}
\includegraphics[angle=-90,width=0.162\textwidth,totalheight=0.32\textwidth]{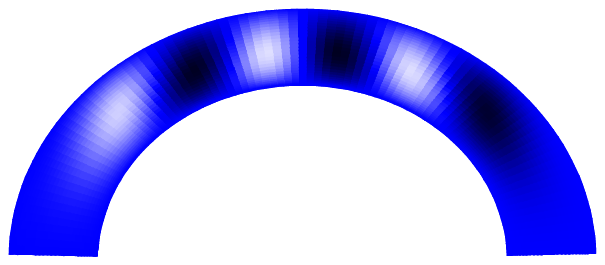}
\includegraphics[angle=-90,width=0.162\textwidth,totalheight=0.32\textwidth]{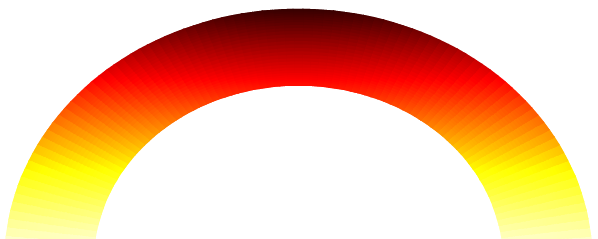}
\includegraphics[angle=-90,width=0.162\textwidth,totalheight=0.32\textwidth]{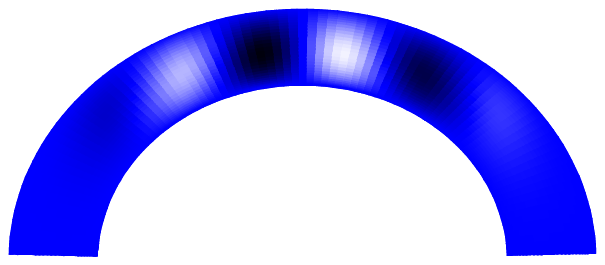}
\includegraphics[angle=-90,width=0.162\textwidth,totalheight=0.32\textwidth]{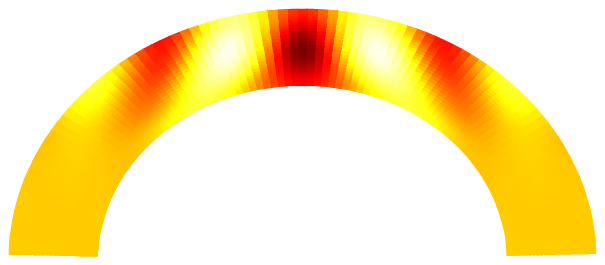}
\includegraphics[angle=-90,width=0.162\textwidth,totalheight=0.32\textwidth]{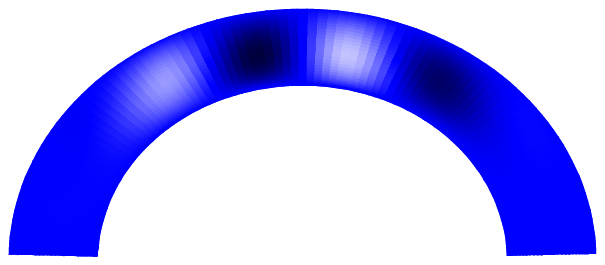}
\includegraphics[angle=-90,width=0.162\textwidth,totalheight=0.32\textwidth]{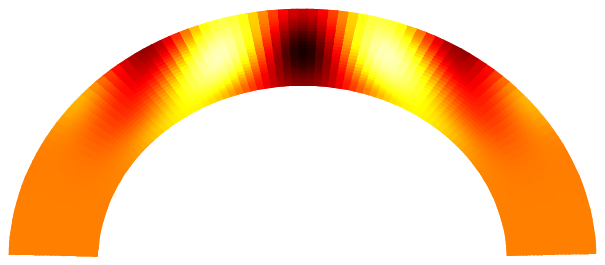}
\par (a)\hspace{5.4cm}(b)\hspace{5.4cm}(c)
\caption{ Vertical cross-sections of the magnetic field (black-blue) and the velocity field (red-yellow).
         (a) Toroidal fields for $R_B=1$ and $S=0$, pure SHI.
         (b) Toroidal fields for $R_B=0$ and $S=70$, pure MRI.
         (c) Toroidal fields for $R_B=1$ and $S=70$, Hall-MRI.
        }\label{fig:grA0tor}
\end{figure*}

\begin{figure}
	\includegraphics[clip,width=0.48\textwidth]{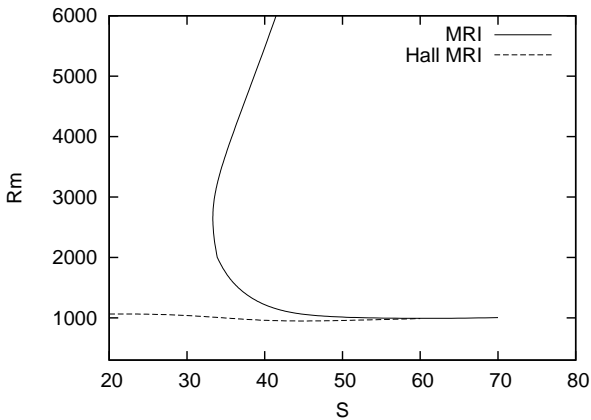}

	\caption{Critical magnetic Reynolds number for the pure MRI and the
        Hall-MRI for the $m=1$ mode for which the magnetic field is
        symmetric with respect to the equator. The Lundquist number $S$ is
        varied, while the Hall parameter $R_B=1$ is fixed. 
	} \label{fig:sm_S_S1_TK} 

\end{figure}

\subsection{Field topology}

Figure~\ref{fig:grA0tor} illustrates
the topology of the toroidal component of the most unstable A0 mode for
${\mrn=3000}$ in the situations when the pure SHI, or the pure MRI or both
mechanisms responsible for the field amplification are at work. The rotation axis and the background magnetic field are parallel here. The
formation of zones which grow in number with the increase of the magnetic Reynolds number (here ${\mrn=3000}$)
is already known from \citet{korue2011}.  Another interesting feature,
evident from comparing Figs.~\ref{fig:grA0tor}a and
\ref{fig:grA0tor}b, is that the corresponding sections of the two
wound-up toroidal field plots have the opposite signs. This effect is a
consequence of the entirely different manner in which the MRI and the
SHI generate the radial field component (which is then wound up into the
toroidal field by the shear).  Both mechanisms acting together will
interfere and decrease the efficiency of the amplification. 

Additionally, since the Lorentz force is invariant under the
transformation ${B_{0} \rightarrow -B_{0} }$, while the Hall term
changes sign, after the transformation the Hall term will produce the
toroidal field in the opposite direction. The stability limit is increased correspondingly for positive $\beta$ in Fig.~\ref{fig:staAS070} as compared to the $\beta=0$ stability limit. This is why both processes
interfere constructively in case of $B_{0}<0$ (anti-parallel to the rotation axis) and small enough $\beta$. The stability limit goes below the $\beta=0$ limit in the left-hand side of Fig.~\ref{fig:staAS070}.
These results are comparable to the local linear stability analysis of
\citet{bate2001}.

%
%
%
%

\begin{figure}
	\includegraphics[clip,width=0.48\textwidth]{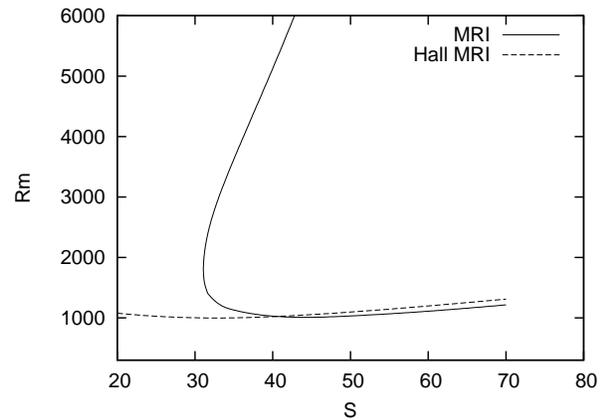}

	\caption{Critical magnetic Reynolds number as in Fig.~\ref{fig:sm_S_S1_TK} 
        but for the $m=1$ mode which is anti-symmetric with respect to the equator.
	} \label{fig:sm_S_A1_TK} 

\end{figure}

\section{Non-axisymmetry}
The marginal stability of an $m=1$ perturbation was also studied in this
setup. The critical magnetic Reynolds numbers for pure MRI and the
Hall-MRI are shown in Figs.~\ref{fig:sm_S_S1_TK} and \ref{fig:sm_S_A1_TK}
for the S1 and A1 modes, respectively, as a function of $S$.
A typical feature for the stability of non-axisymmetric perturbations
in differentially rotating system is the presence of an upper
limit for the Reynolds number (see also \citealt{kiru2010} for similar behaviour in disks). 
At $S=40$, for example, there is an instability window for S1 between 
$1130<{\rm Rm}<5410$. There is no instability for ${S<33}$, while 
the system is still unstable against $m=0$ perturbations. For small fields,
the MRI only produces axisymmetric field geometries. The A1 mode is 
unstable for ${1030<{\rm Rm}<5130}$ at ${S=40}$ and entirely stable 
against pure MRI for $S<30.5$.

The situation is again changed by the presence of the Hall effect. 
Figure~\ref{fig:sm_S_S1_TK} shows the critical Rm for ${R_B=1}$. There
is practically no longer any dependence on $S$, with critical Rm
always being near 1000. The critical values are usually slightly
above the marginal Rm for the $m=0$ modes, but can be lower for
large  Lundquist numbers $S$ (e.g.\ compare $S=70$ in Fig.~\ref{fig:sm_S_S1_TK}
with $\beta=0.014$ in Fig.~\ref{fig:staAS070}). There is obviously more
non-axisymmetry possible in this differentially rotating system if 
the Hall effect is present.

\begin{figure}
	\includegraphics[clip,width=0.48\textwidth]{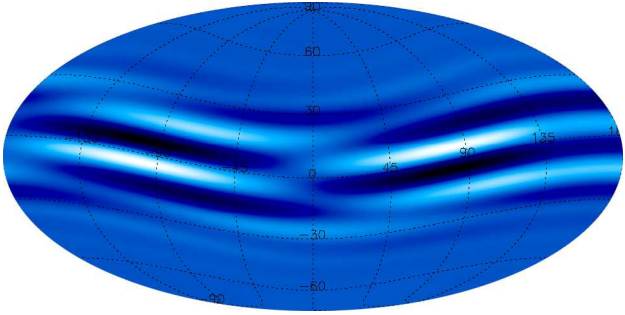}

	\caption{ Surface map of the radial component of the magnetic field for an the case of an
        unstable A1 mode at ${{\rm Rm}= 5000}$, ${S=35}$, and ${\beta=1}$. Light areas represent
        $b_r > 0$, while dark areas are $b_r < 0$.
	} \label{fig:field_rb1_rm5000_S35_pm1_A1_jpg} 

\end{figure}

A surface plot of the magnetic field pattern emerging from an unstable
A1 mode is shown in Fig.~\ref{fig:field_rb1_rm5000_S35_pm1_A1_jpg}.
Like for the axisymmetric modes, the eigenmode is concentrated near
the equatorial plane. There is no indication for distinct magnetic poles
as one would expect for a tilted dipole. The $m=1$ instability will
probably not directly serve as an explanation for the non-axisymmetric
magnetic fields of pulsars.

\section{Discussion}

\begin{figure}

	\includegraphics[clip,width=0.48\textwidth]{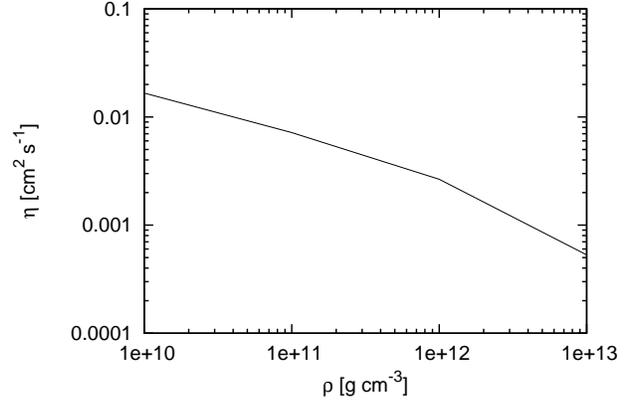}

	\caption{Microphysical magnetic diffusivity as a function of density for a temperature of $10^{10}\ \mathrm{K}$. }
	
	\label{fig:etarho} 

\par\end{figure}

\begin{figure}

	\includegraphics[clip,width=0.48\textwidth]{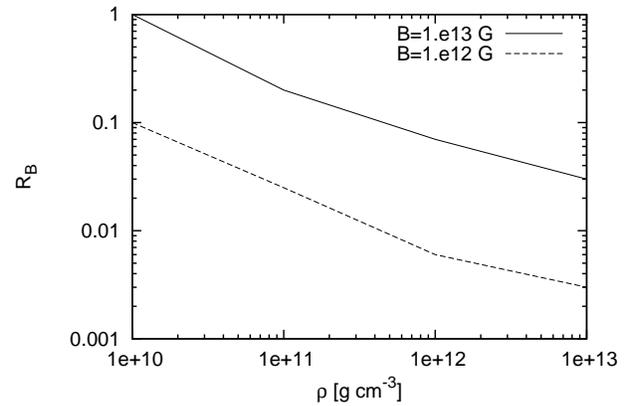}

	\caption{The Hall parameter as a function of density for a temperature of $10^{10}\ \mathrm{K}$. }
	
	\label{fig:rbrho} 

\par\end{figure}

The physical model of PNS can be related to the stability map in Fig.
\ref{fig:staAS0} using the code developed by \citet{po2008} for
calculating the electrical conductivity tensor of dense plasmas, by
taking into account the electron-ion interaction.  However, at 
${T \gtrsim 10^{10}}\ \mathrm{K}$, the plasma of the protoneutron star envelope
contains a distribution of fully ionised atoms, free neutrons, protons
and electrons \citep{hapo2007}. An exact calculation of electrical
conductivities, which are necessary for reconstructing the evolution of
the magnetic field, would need to include the interaction between all
the different particles. This manuscripts considers only the
electron-ion contribution calculated by the code of \citet{po2008}.  The
values of the magnetic diffusivity and the Hall parameter used in this
manuscript should be taken as order-of-magnitude estimates (see  Figs.
\ref{fig:etarho} and \ref{fig:rbrho}).

According to  Fig. \ref{fig:rbrho}, the Hall parameter and, therefore, the
Hall effect vanish in deeper layers of a PNS's envelope even for a moderate
magnetic field of $10^{12}\ {\rm G}$. Only the outer layers with densities in
the range of $\rho\sim 10^{10}$--$10^{11}\, {\rm
g}\,\mbox{cm}^{-3}$ are characterised by the $R_B$ closer to unity.  

If the flow is laminar, $\mrn$ and $S$ are orders of magnitude larger
than the examined parameter space in the stability map of  Fig.
\ref{fig:staAS0}, due to the low diffusivity (see Fig.~\ref{fig:etarho}). If,
however, the flow becomes turbulent, the turbulent diffusivity is much
larger \citep{narebopa2008}, leading to much lower $\mrn$ (${\sim} 10^{4}$, as
noted by \citealt{korue2011}, and which are doable numerically), but still above the threshold for
the onset of the Hall-MRI. An interesting feature of turbulent flows is
that $S$ is close to unity for densities larger than or equal to 
$10^{13}\ {\rm g}\,\mbox{cm}^{-3}$.  These regions will be purely SHI unstable,
because the MRI is inhibited by diffusion.

Therefore, the Hall-MRI instabilities shall set in after the seed field
of the progenitor is sufficiently amplified. They would probably first
destabilise the outer regions, perhaps spreading inward provided that
the fields in the interior become sufficiently strong (because the Hall
parameter $R_B$ scales linearly with the magnetic field).  If the flow 
is already turbulent for other reasons, even the pure SHI may be encountered 
in  regions where the Lundquist number is small enough.

Given that we have shown the growth rates of the instability to depend
on a general orientation of the magnetic field, those multipolar field
components which have the opposite directions in different hemispheres, e.g. a quadrupole or a tilted dipole,
will experience different growth. The fields with dominant negative
orientation would be strongly restricted in their growth. This may have
interesting implications for observable properties of the field, with
one hemisphere having different field strength and patterns than the
other.

The nonlinear effects related to this process are the self-coupling of
the magnetic field through the quadratic Hall term, as well as the
nonlinear nature of the flow/field interaction where Maxwell stresses redistribute the angular momentum in the fluid, braking the
differential rotation on the time-scale comparable to
the Alfv\'{e}n time (shorter than any time-scale related to viscosity
effects).
These effects will need to be taken into account in order to obtain the
estimates of the final amplitude and topology of the magnetic field
available through the Hall-MRI. The proximity of the excitation thresholds
of axisymmetric and non-axisymmetric modes is another interesting outcome
of this paper; the nonlinear coupling of the modes may lead to different 
classes of surface field topologies with implications for the observability 
of neutron stars.

\end{document}